# High-power, continuous-wave, tunable mid-IR, higher-order vortex beam optical parametric oscillator

A. Aadhi,* Varun Sharma, and G. K. Samanta

Photonic Sciences Lab, Physical Research Laboratory, Navarangpura, Ahmedabad 380009, Gujarat, India
*Corresponding author: aadhi@prl.res.in

**We report on a novel experimental scheme to generate continuous-wave (cw), high power, and higher-order optical vortices tunable across mid-IR wavelength range. Using cw, two-crystal, singly resonant optical parametric oscillator (T-SRO) and pumping one of the crystals with Gaussian beam and the other crystal with optical vortices of orders, $l_p$ = 1 to 6, we have directly transferred the vortices at near-IR to the mid-IR wavelength range. The idler vortices of orders, $l_i$ = 1 to 6, are tunable across 2276–3576 nm with a maximum output power of 6.8 W at order of, $l_i$ = 1, for the pump power of 25 W corresponding to a near-IR vortex to mid-IR vortex conversion efficiency as high as 27.2%. Unlike the SROs generating optical vortices restricted to lower orders due to the elevated operation threshold with pump vortex orders, here, the coherent energy coupling between the resonant signals of the crystals of T-SRO facilitates the transfer of pump vortex of any order to the idler wavelength without stringent operation threshold condition. The generic experimental scheme can be used in any wavelength range across the electromagnetic spectrum and in all time scales from cw to ultrafast regime.**

Optical vortices having phase singularities or the topological defects carry doughnut shaped intensity pattern. The characteristic azimuthal (helical) phase variation of the vortices is given as, $exp(\pm il\varphi)$, where, $\varphi$ is the azimuthal angle and the integer, $l$, known as the topological charge or vortex order, indicates that the vortex beam carry orbital angular momentum (OAM) of $l\hbar$ per photon. Coherent sources of optical vortices at different wavelengths across the electromagnetic spectrum are of great interest for a variety of applications in science and technology including the increase of channel capacity of information carrier [1], generation of high-dimensional quantum state [2], improvement in the resolution of optical microscopy [3] and optical tweezing [4]. Such applications continue to motivate to explore different mechanisms to generate optical vortices, however, the mode conversion of Gaussian laser beam using different mode converters, including computer generated hologram in spatial light modulators (SLM) [5], spiral phase plates (SPP) [6], cylindrical lens [7] and q-plates [8], remained the ultimate choice. However, all these mode converters suffer from at least one of the common drawbacks such as low damage threshold, low wavelength coverage and high cost.

In recent times, a significant efforts have been made to generate optical vortices using conventional techniques and transferred to new wavelengths at limited or no wavelength tunability through nonlinear frequency conversion processes such as sum frequency generation (SFG) [9], second harmonic generation (SHG) [10] and optical parametric generation (OPG) [11]. Similarly, the intrinsic advantages of the optical parametric oscillators (OPOs) in terms of high output power and wide wavelength coverage [12] have been explored in recent times to generate optical vortices tunable over wide wavelengths across the electromagnetic spectrum in different time scales (cw to ultrafast) [13-15]. However, the elevated operation threshold due to the lower parametric gain of vortex pump beams especially at cw time scale has restricted the OPOs producing lower order optical vortices [15]. To avoid such stringent threshold requirement, one can, in principle, use optical parametric amplification (OPA) process and transfer the OAM of the pump beam and or the seed beam to the generated beam. Here, we report, first time to the best of our knowledge, a novel experimental scheme using two-crystal based cw, singly resonant OPO (T-SRO) [16-17] producing high power and higher order optical vortex beams tunable in the mid-IR wavelength range. Using two identical MgO doped periodically poled lithium niobate (MgO:PPLN) crystals independently pumped with Gaussian beam and optical vortex beams at 1064 nm, we have transferred the pump vortex of orders, $l_p$ = 1-6 to the idler vortices of orders, $l_i$ = 1-6, tunable across 2276-3576 nm while restricting the resonant signal beam in fundamental (Gaussian, $l$ = 0) cavity mode.

The schematic of the experimental setup is shown in Fig. 1. A cw, single-frequency (line-width of <100 KHz) Yb-fiber laser (IPG Photonics, YLR-50-1064-LP-SF) at 1064 nm delivering a maximum output power of 50 W, is divided into two beams (PB1, PB2) to pump the T-SRO. The power of each beam to the T-SRO is controlled using standard power attenuator (not shown here) comprising with a half-wave plate, ($\lambda$/2), and polarizing beam splitter cube [16]. The $\lambda$/2 plates in each of the pump beams, PB1, PB2, are used to control the polarization of the pump beams with respect to crystal orientation for optimum phase-matching. Two spiral phase plates, SPPs, of winding number corresponding to vortex orders, $l$=1, and 2, and a vortex doubling setup [18] (not shown here) are used to convert the Gaussian pump beam, PB2, into vortices of orders, $l_{p2}$ = 1-6, while the pump beam, PB1, remains in Gaussian spatial profile ($l_{p1}$=0). Two lenses, L1 and L2, of equal focal length, $f$ = 150 mm, are used to focus the pump beams, PB1 and PB2, respectively, at the centre of the nonlinear crystals. The Gaussian pump beam waist radius at the centre of the crystal, C1, is estimated to be ~84 μm. The T-SRO is designed in a compact four plano-concave mirrors, M1-4, cavity with two crystals at two foci of the cavity. All the curve mirrors, M1-4 (radius of curvature, $r$ =150 mm), have high reflectivity (R>99%) at signal (1400-2000 nm) and high transmission (T>95%) at idler (2200-4000 nm) and pump wavelengths ensuring singly resonance condition. Two identical crystals, C1 and C2, of 50-mm-long and 8 x 1 mm² in aperture multi-grating, MgO:PPLN crystals with periods, $\Lambda$ = 28.5-31.5 μm at an increment of 0.5 μm, placed at two foci of the T-SRO cavity respectively in between M1 and M2, and M3 and M4, respectively, are used as nonlinear crystals. Both the crystals are anti-reflective coated for the pump, signal and idler wavelengths. The crystals are housed in two separate temperature controlled oven whose temperature can be varied up to 200°C with a stability of ±0.1°C. A temperature off-set of ~9-20°C, depending upon the grating periods, is observed between the two crystal ovens while generating signals of same wavelength. The idler beams generated from the crystals, C1 and C2, are extracted from corresponding residual pump beams using wavelength separator, S. The total optical length of the cavity in presence of both the crystal is measured to be 760 mm.

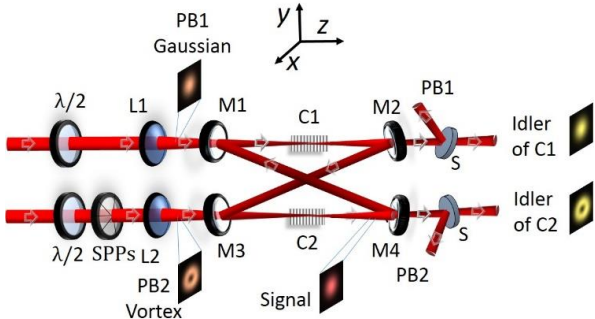

Fig. 1. Schematic of the experimental setup. λ/2, half-wave plate; L1-L2, lenses; SPPs, spiral phase plates; C1-C2, MgO:PPLN crystals; M1-4, cavity mirrors; S, wavelength separator. (Inset) Typical intensity distribution of the interacting beams in the experiment.

To generate higher order optical vortices, we first verified the coherent energy coupling between the resonant signals of both the crystals. Maintaining same grating period, $\Lambda_{C1} = \Lambda_{C2}$ = 31.5 $\mu m$, and temperature, $T_{C1} = T_{C2}$ = 190°C in both crystals, we have pumped the crystal, C1, with Gaussian beam ($l_{p1}$ = 0), PB1, of power, $P_G$ = 5 W, and the crystal, C2, with vortex beam, PB2, of order, $l_{p2}$ = 1, and power, $P_V$ = 1 W. Since the pump power, $P_G$ = 5 W, is higher than the operation threshold (~2 - 3 W) of MgO:PPLN crystal based SRO [19] we observed idler (signal) radiation of wavelength 2276 nm (1998 nm) from crystal C1 and no output from the crystal, C2. Keeping everything fixed, we gradually changed the temperature of crystal, C2, and found idler output of wavelength 2276 nm from crystal, C2 at temperature, $T_{C2}$ = 200°C. Although, the pump power to crystal C2 is lower than the operation threshold of vortex pumped SRO [15], the generation of idler of wavelength 2276 nm from the crystal, C2, can be attributed to the coherent energy coupling between the resonant signals at wavelength 1998 nm of crystals, C1 and C2, under identical phase-matching condition. The off-set in the crystal temperatures, $T_{C1}$ = 190°C and $T_{C2}$ = 200°C, can be attributed the differences in the heating parameters of the ovens and the imperfection in the grating periods of the crystals. Using the coherent energy coupling configuration, we have changed the orders of the pump vortex beam, PB2, and studied the output beams with the results shown in Fig. 2. The first column, (a-c), of Fig. 2, shows the doughnut intensity distribution with increasing dark core size with the increase of pump vortex orders, $l_{p2}$ = 1, 3, and 6, respectively. Using the tilted lens technique [20] and counting the number of lobes, as shown in second column, (d-f), of Fig. 2, we confirmed the pump vortex orders to be, $l_{p2}$ = 1, 3, and 6, respectively. The third column, (g-i), of Fig. 2, shows the intensity distribution of the idler beam, generated from crystal, C1, in Gaussian ($l_{i1}$ = 0) spatial profile without any phase singularity. However, the idler beams generated from crystal, C2, have doughnut shaped intensity distribution, as shown in fourth column, (j-l), of Fig. 2, with increasing dark core size. The fifth column, (m-o), of Fig. 2, shows the lobe structure of the idler vortices passed through the tilted lens and confirms the order of the idler vortices to be, $l_{i2}$ = 1, 3, and 6, respectively, same as the pump orders, $l_{p2}$ = 1, 3 and 6. Using the leakage beam of the T-SRO we observed the spatial profile of the resonant signal, as shown in sixth column, (p-r), of Fig. 2, in Gaussian spatial profile for all pump vortex orders. As expected, the SRO based on crystal, C1, operated close to its operation threshold produces signal and idler both in Gaussian ($l_s = l_{i1}$ = 0) spatial profile for Gaussian pump beam ($l_{p1}$ = 0). However, due to the lowest cavity loss at the fundamental cavity mode, the SRO cavity restricts the resonant signal in Gaussian mode, ($l_s$ = 0), and transfer the pump OAM ($l_{p2}$) in crystal, C2, to the idler ($l_{i2} = l_{p2}$) owing to the OAM conservation, $l_{p2} = l_s + l_{i2}$, in nonlinear processes. The coherent energy coupling between the resonant signals of both the crystals helps the transfer of pump vortex of any order to the idler wavelength without any operation threshold. Although, we have restricted our study to the vortices of orders up to, $l$ = 6, due to the limited resources available in our lab, using the current scheme, one can, in principle, generate high power and higher order vortices in the mid-IR. However, the lower parametric gain due to the lower overlapping integral between the Gaussian signal beam with the higher order pump vortices can be the major concern for the generation of higher order optical vortices at high output power.

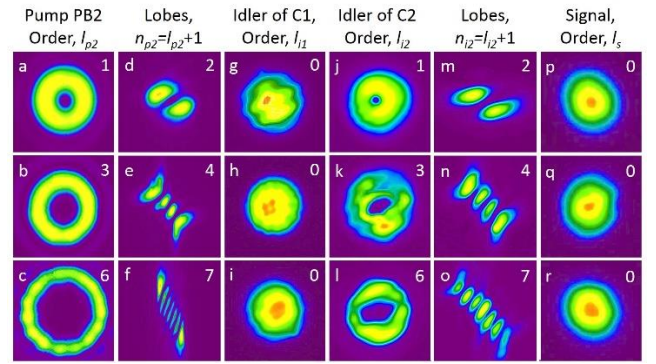

Fig. 2. (a-c) Far field intensity distribution, and (d-f), corresponding lobe structures of the pump vortex beam, PB2, of orders, $l_{p2}$=1, 3, and 6. (g-i), Intensity distribution of the idler generated by crystal, C1, for Gaussian ($l_{p1}$ =0) pump beam. (j-l) Intensity distribution, and (m-o), corresponding lobe structures of the idler beam of crystal, C2, generated by the pump vortex beam of orders, $l_{p2}$ =1, 3, and 6. (p-r) Intensity distribution of the resonating signal beam generated by both vortex and Gaussian pump beams. The pump, signal and idler wavelengths are 1064 nm, 1998 nm and 2276 nm, respectively.

To study the presence of vortex profile across the tuning range, we pumped the crystals, C1 and C2, with Gaussian beam, $l_{p1} = 0$, and vortex beam of order, $l_{p2} = 1$, respectively, at same powers, $P_V = P_G = 5$ W. Adjusting the temperature and grating periods of the crystals to maintain coherent energy coupling between the resonant signals, we have measured the intensity distribution of the idler beams and the leakage signal beam of the T-SRO across the tuning range. The results are shown in Fig. 3. As evident from the first column, (a-d), of Fig. 3, the resonant signal tunable across 1972-1559 nm maintains a Gaussian spatial profile, $l_s = 0$. The corresponding idler generated by the crystal, C1, tunable across 2311-3351 nm, as shown in second column, (e-h), of Fig. 2, has Gaussian spatial profile, $l_{i1} = 0$. However, the idler generated by the crystal, C2, for pump vortex of order, $l_{p2} = 1$, as shown in third column, (i-l), of Fig. 3, carries doughnut intensity distribution resembling vortex beam across the tuning range of 2311-3351 nm. To determine the vortex order, we have passed the idler beam through a tilted lens [20] to split the beam into its characteristic lobes with the results shown in fourth column, (m-p), of Fig. 3. The two-lobe intensity structure of the idler beam, as evident from fourth column, (m-p), of Fig. 3, confirms the idler vortex order of crystal, C2, to be, $l_{i2} = 1$, and thus, verifying the generation of idler vortex across the tuning range across 2311 - 3351 nm.

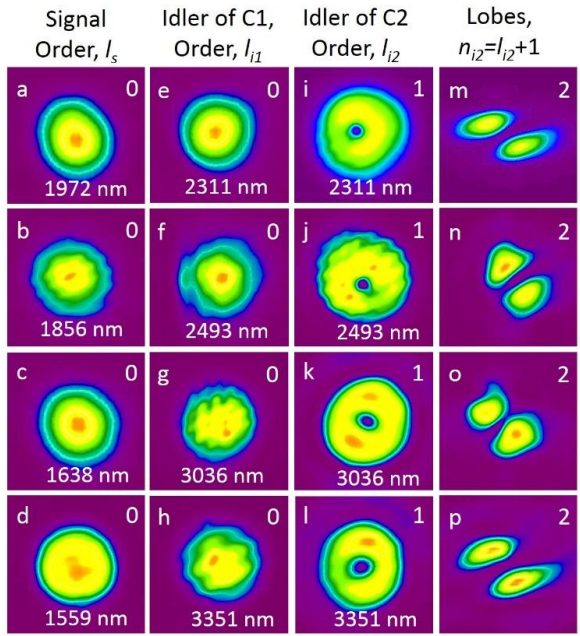

Fig. 3. Intensity distribution of (a-d) resonating signals, and the corresponding, (e-h), idler beam generated by crystal, C1, under Gaussian beam pumping. (i-l) Intensity distribution, and the (m-p) lobe structure of the idler beam generated by crystal, C2, for vortex beam pumping across the tuning range of the T-SRO.

Further, we have studied the performance of the T-SRO in terms of output power of the Gaussian and vortex idler beams, and vortex pump depletion across the tuning range, the maximum power of the vortex beam with orders, and the power scaling characteristics. The results are shown in Fig. 4. Pumping the crystals, C1 and C2, with Gaussian and vortex beams of equal powers, $P_G = P_V = 15$ W, respectively, and using two grating periods, $\Lambda = 30$ $\mu$m and 31.5 $\mu$m, and varying the crystal temperatures across 30-200°C under coherent energy coupling of the resonant signals, we have measured the output power of both the idler beams with the results shown in Fig. 4(a). As evident from Fig. 4(a), the output power of the vortex (Gaussian) idler beam of order, $l_{i2} = 1$ ($l_{i1} = 0$) varies from 4.2 W (4.8 W) at 2272 nm to 1.62 W (2 W) at 3574 nm with a maximum vortex-vortex (Gaussian-Gaussian) conversion efficiency of 28% (32%) at 2272 nm. The relatively higher idler power in Gaussian spatial distribution can be attributed to the higher nonlinear gain in crystal, C1, arising from the higher mode overlapping of the Gaussian resonant signal beam with the Gaussian pump beam than that with vortex pump beam in crystal, C2. The idler vortex (Gaussian) beam has output power in access of 1.25 W (2 W) throughout tuning range of 1302 nm across 2272 - 3574 nm. We have also recorded the depletion of vortex pump beam, as shown in Fig. (b), to be > 40% across the tuning range with maximum pump depletion of ~66 % at the idler wavelength of 2276 nm.

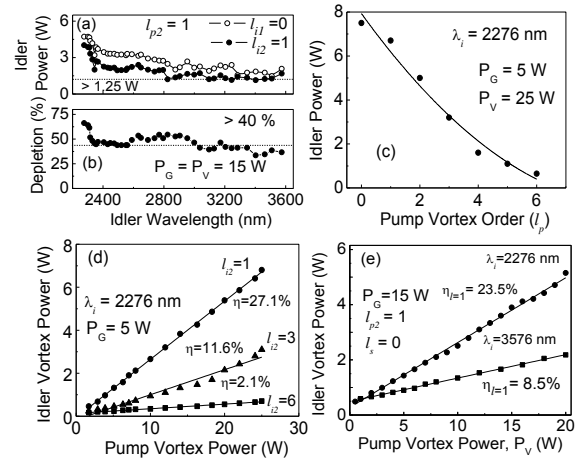

Fig. 4. Variation of (a) vortex and Gaussian idler beam power, and the (b) pump depletion across the tuning range of T-SRO. (c) Dependence of idler vortex power on pump vortex orders. (d) Variation of idler vortex power of orders, $l_{i2}$ =1, 3, and 6, as a function of pump vortex powers. (e) Dependence of output power of vortex idler beam of order, $l_{i2}$ =1, on the pump power at two different wavelengths, 2276 nm and 3576 nm, across the tuning range.

Pumping the crystal, C1, with Gaussian pump of power, $P_G = 5$ W, we have changed the order of the pump vortex from, $l_{p2} = 0$ (Gaussian) to $l_{p2} = 6$ at constant power of, $P_V = 25$ W, and measured the idler power from the crystal, C2, at 2276 nm with the results shown in Fig. 4(c). While the order of the idler vortex beam is changing with the order of the pump beam, we observe, as evident from Fig. 4(c), a continuous decrease in the output power from 7.5 W at vortex order of, $l_{i2} = 0$ (Gaussian beam), to 0.64 W for the vortex order of, $l_{i2} = 6$. Such decrease can be attributed to the decrease in the nonlinear gain due to the increase in the beam area and beam divergence of the pump vortices with its order [18]. To study the power scaling characteristics of the T-SRO, we have operated the source at idler wavelength of 2276 nm with the Gaussian beam of power, $P_G = 5$ W. Varying the power of the pump vortex beams of orders, $l_{p2} = 1$, 3, and 6, we have recorded the idler vortex power with the results shown in Fig. 4(d). As evident from Fig. 4(d), the idler vortex power of orders, $l_{i2} = 1$, 3 and 6, increases linear to the pump power at a slope efficiency of 27.1%, 11.6% and 2.1%, respectively, without any sign of saturation. The maximum idler power is measured to be 6.8 W, 3.1 W, and 0.7 W of orders, $l_{i2} = 1$, 3, and 6, respectively, at constant pump power of 25 W. Although, one can, in principle, further increase the vortex power with the increase of pump power, however, we have restricted our study to the pump power of 25 W to avoid possible thermal damage in the nonlinear crystal. Additionally, we have also studied the power scaling of the vortex source at two different wavelengths, 2276 and 3576 nm, across the tuning range. Pumping the T-SRO with Gaussian beam of power of 15 W, we have measured the vortex idler with the increase of pump power of vortex order, $l_{p2}$ = 1, with the results shown in Fig. 4(e). As evident from Fig. 4 (e), the source generates vortex idler beam of order, $l_{i2}$ = 1, at a slope efficiency of 23.5% and 8.5% and maximum output power of 5.2 W and 2.2 W at wavelengths of 2276 and 3576 nm, respectively, for the pump power of 25 W. The reduction in the slope efficiency and maximum output power of the vortex idler beam at 3576 nm can be attributed to the lower parametric gain of the nonlinear crystal for the phase matching wavelength away from the degeneracy.

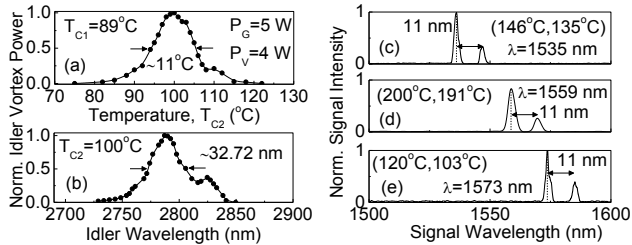

Fig. 5. Determination of (a) temperature, and (b) spectral acceptance bandwidth of a 50 mm long MgO:PPLN crystal. Appearance of Raman peak at different combinations of crystal temperatures, ($T_{C1}$, $T_{C2}$), of (c) (146°C, 135°C), (d) (200°C, 191°C) and (e) (120°C, 103°C) resulting signal wavelengths of 1535 nm, 1559 nm and 1573 nm, respectively.

We have also measured the temperature and spectral acceptance bandwidth of 50 mm long MgO:PPLN crystal. We pumped the crystal, C1, at grating period and temperature of, $\Lambda$ = 31.5 $\mu$m and $T_{C1}$ = 89°C respectively, with Gaussian beam power of 5 W to produce signal (idler) wavelength of 1720 nm (2789.7 nm). Keeping the same grating period in crystal, C2, and pumping with vortex beam of power, 4 W, of order, $l_{p2}$ = 1, we have varied the temperature, $T_{C2}$, and measured corresponding idler power. The results are shown in Fig. 5(a). Since the pump vortex power (4 W) is far below the threshold of the T-SRO, we did not observe any idler output of the crystal, C2, for $T_{C2}$ far away from $T_{C1}$. However, with the increase of $T_{C2}$ from 80°C to 120°C we observe the vortex idler power to increase and finally decrease to zero showing a clear peak near the crystal temperature, $T_{C2}$=100°C. Considering the temperature off-set between two ovens, we find, as expected, the peak in the idler power when both the crystals maintain same phase-matching condition. Using the results of Fig. 5(a), we have measured the temperature acceptance bandwidth (full-width at half maximum, FWHM) of a 50 mm long MgO:PPLN crystal to be ~11°C. Similarly, to measure the spectral acceptance bandwidth, we adjusted $T_{C2}$ = 100°C and varied $T_{C1}$ to change the wavelength of the resonating signal and measured the wavelength and output power of the vortex idler. The results are shown in Fig. 5(b). As evident from Fig. 5(b), for the change of $T_{C1}$ across 70 - 110°C, the crystal, C2, produces idler across 2745-2850 nm at a FWHM bandwidth of 32.7 nm centred at 2789.7 nm for $T_{C1}$ = 89°C. Using the pump wavelength of 1064 nm and the idler wavelengths of Fig. 5(b), we have calculated the spectral acceptance bandwidth of a 50 mm long MgO:PPLN crystal to be ~12.56 nm, centred at 1720 nm.

To confirm coherent energy coupling between the resonant signals, we have pumped the crystals, C1, and C2, with ($P_G$ = $P_V$ = 5 W) and measured the signal spectrum at different combinations of crystal temperatures, ($T_{C1}$, $T_{C2}$), to maintain same phase-matching condition. The results are shown in Fig. 5(c-e). As evident from Fig. 5(c-e), the T-SRO produces signal wavelengths of 1535 nm, 1559 nm and 1579 nm across the tuning range for the crystal temperatures, ($T_{C1}$, $T_{C2}$), of (146°C, 135°C), (200°C, 191°C) and (120°C, 103°C), respectively. However, it is interesting to note that for all signal wavelengths we observe an additional spectral line red shifted by ~11 nm (~45 cm$^{-1}$).

Such additional spectral line with constant wavelength shift can be attributed to the Raman effect in MgO:PPLN crystal arising from the high intra-cavity power and long interaction length of the nonlinear crystal due to the coherent energy coupling between the resonating signals. Using scanning confocal Fabry–Perot interferometers (free spectral range, 10 GHz and finesse 150) we have measured the instantaneous line-width of the signal and idler radiation to be ~15 MHz and ~205 MHz, respectively across the tuning range and independent to the vortex orders. We also observed the idler vortex beam ($l_{i2}$ = 1) at 2276 nm, under the free running condition, to exhibit *rms* power stability better than 6% over 1 hour, which can further be improved by active cavity stabilization and thermal isolation of the system from the laboratory environment.

In conclusion, we have demonstrated a novel experimental scheme to generate high power and higher order optical vortices in the mid-IR wavelength range. Using two nonlinear crystals of same characteristics coupled in a single OPO cavity and pumping independently with Gaussian and vortex beams, we transferred the cw pump vortices of orders, $l_p$ = 1-6, at 1064 nm into optical vortices of same orders in the mid-IR wavelength range tunable across 2276 - 3576 nm. Using 25 W of pump vortex power we have generated idler vortex of order, $l$ = 1, of power as high as 6.8 W at 2276 nm. Using two crystals inside the cavity, we have also measured the temperature and spectral acceptance bandwidth of 50 mm long MgO:PPLN crystal for OPO operation. The generic experimental configuration can be used for any vortex orders and different structure beams at any wavelength across the electromagnetic spectrum in any time scales, cw to ultrafast.

**References**

1. J. Wang, J. Y. Yang, I. M. Fazal, N. Ahmed, Y. Yan, H. Huang, Y. Ren, Y. Yue, S. Dolinar, M. Tur, and A. E. Willner, Nature Photon. **6**, 488 (2012).
2. A. Mair, A. Vaziri, G. Weihs, and A. Zeilinger, Nature **412**, 313 (2001).
3. S. W. Hell, Nat. Biotechnol. **21**, 1347 (2003).
4. D. G. Grier, Nature **424**, 810 (2003).
5. N. R. Heckenberg, R. McDuff, C. P. Smith, and A. G. White, Opt. Lett. **17**, 221 (1992).
6. S. S. R. Oemrawsingh, J. A. W. Van Houwelingen, E. R. Eliel, J. P. Woerdman, E. J. K. Verstegen, and J. G. Kloosterboer, App. Optics, **43**, 688 (2004).
7. M. W. Beijersbergen, L. Allen, H. Van der Veen, and J. P. Woerdman, Opt. Comm. **96**, 123 (1993).
8. L. Marrucci, C. Manzo, and D. Paparo, Phys. Rev. Lett. **96**, 163905 (2006).
9. A. Beržanskis, A. Matijošius, A. Piskarskas, V. Smilgevičius, and A. Stabinis, Opt. Comm. **150**, 372 (1998).
10. N. Apurv Chaitanya, S. Chaitanya Kumar, K. Devi. G. K. Samanta, and M. Ebrahim-Zadeh, Opt. Lett. **41**, 2715 (2016).
11. A. Aadhi, and G. K. Samanta, J. Optics **20**, 01LT01 (2017).
12. M. Ebrahim-Zadeh, S. Chaitanya Kumar, A. Esteban-Martin, and G. K. Samanta, IEEE Photonics J. **5**, 0700105 (2013).
13. A. Abulikemu, T. Yusufu, R. Mamuti, K. Miyamoto, and T. Omatsu, Opt. Express **23**, 18338 (2015).
14. A. Aadhi, G. K. Samanta, S. Chaitanya Kumar, and M. Ebrahim-Zadeh, Optica **4**, 349 (2017).
15. A. Aadhi, V. Sharma, R. P. Singh, and G. K. Samanta, Opt. Lett. **42**, 3674 (2017).
16. G. K. Samanta, A. Aadhi, and M. Ebrahim-Zadeh, Opt. Express, **21**, 9520 (2013).
17. G. K. Samanta and M. Ebrahim-Zadeh, Opt. Lett. **36**, 3033 (2011).
18. N. Apurv Chaitanya, A. Aadhi, M. V. Jabir, and G. K. Samanta, Opt. Lett. **40**, 2614 (2015).
19. S. Chaitanya Kumar, R. Das, G. K. Samanta, and M. Ebrahim-Zadeh, Appl. Phys. B **102**, 31 (2011).
20. Pravin Vaity, J. Banerji, and R. P. Singh, Phys. Lett. A **377**, 1154 (2013).

**References (with titles)**

1. J. Wang, J. Y. Yang, I. M. Fazal, N. Ahmed, Y. Yan, H. Huang, Y. Ren, Y. Yue, S. Dolinar, M. Tur, and A. E. Willner, "Terabit free-space data transmission employing orbital angular momentum multiplexing," Nature Photon. **6**, 488 (2012).
2. A. Mair, A. Vaziri, G. Weihs, and A. Zeilinger, "Entanglement of the orbital angular momentum states of photons," Nature **412**, 313 (2001).
3. S. W. Hell, "Toward fluorescence nanoscopy," Nat. Biotechnol. **21**, 1347 (2003).
4. D. G. Grier, "A revolution in optical manipulation," Nature **424**, 810 (2003).
5. N. R. Heckenberg, R. McDuff, C. P. Smith, and A. G. White, "Generation of optical phase singularities by computer-generated holograms," Opt. Lett. **17**, 221 (1992).
6. S. S. R. Oemrawsingh, J. A. W. Van Houwelingen, E. R. Eliel, J. P. Woerdman, E. J. K. Verstegen, and J. G. Kloosterboer, "Production and characterization of spiral phase plates for optical wavelengths," App. Optics, **43**, 688 (2004).
7. M. W. Beijersbergen, L. Allen, H. Van der Veen, and J. P. Woerdman, "Astigmatic laser mode converters and transfer of orbital angular momentum," Opt. Comm. **96**, 123-132 (1993).
8. L. Marrucci, C. Manzo, and D. Paparo, "Optical spin-to-orbital angular momentum conversion in inhomogeneous anisotropic media," Phys. Rev. Lett. **96**, 163905 (2006).
9. A. Beržanskis, A. Matijošius, A. Piskarskas, V. Smilgevičius, and A. Stabinis, "Sum-frequency mixing of optical vortices in nonlinear crystals," Opt. Comm. **150**, 372 (1998).
10. N. Apurv Chaitanya, S. Chaitanya Kumar, K. Devi. G. K. Samanta, and M. Ebrahim-Zadeh, "Ultrafast optical vortex beam generation in the ultraviolet," Opt. Lett. **41**, 2715 (2016).
11. A. Aadhi, and G. K. Samanta, "High-power, high repetition rate, tunable, ultrafast vortex beam in the near-infrared," J. Optics **20**, 01LT01 (2017).
12. M. Ebrahim-Zadeh, S. Chaitanya Kumar, A. Esteban-Martin, and G. K. Samanta, "Breakthroughs in photonics 2012: breakthroughs in optical parametric oscillators," IEEE Photonics J. **5**, 0700105 (2013).
13. A. Abulikemu, T. Yusufu, R. Mamuti, K. Miyamoto, and T. Omatsu, "Widely-tunable vortex output from a singly resonant optical parametric oscillator," Opt. Express **23**, 18338 (2015).
14. A. Aadhi, G. K. Samanta, S. Chaitanya Kumar, and M. Ebrahim-Zadeh, "Controlled switching of orbital angular momentum in an optical parametric oscillator," Optica **4**, 349 (2017).
15. A. Aadhi, V. Sharma, R. P. Singh, and G. K. Samanta, "Continuous-wave, singly resonant parametric oscillator-based mid-infrared optical vortex source," Opt. Lett. **42**, 3674 (2017).
16. G. K. Samanta, A. Aadhi, and M. Ebrahim-Zadeh, "Continuous-wave, two-crystal, singly-resonant optical parametric oscillator: Theory and experiment," Opt. Express, **21**, 9520 (2013).
17. G. K. Samanta and M. Ebrahim-Zadeh, "Dual-wavelength, two-crystal, continuous-wave optical parametric oscillator," Opt. Lett. **36**, 3033-3035 (2011).
18. N. Apurv Chaitanya, A. Aadhi, M. V. Jabir, and G. K. Samanta, "Frequency-doubling characteristics of high-power, ultrafast vortex beams," Opt. Lett. **40**, 2614-2617 (2015).
19. S. Chaitanya Kumar, R. Das, G. K. Samanta, and M. Ebrahim-Zadeh, "Optimally-output-coupled, 17.5 W, fiber-laser-pumped continuous-wave optical parametric oscillator," Appl. Phys. B **102**, 31-35 (2011).
20. Pravin Vaity, J. Banerji, and R. P. Singh, "Measuring the topological charge of an optical vortex by using a tilted convex lens," Phys. Lett. A **377**, 1154 (2013).